\documentclass[pra,aps]{revtex4}

\usepackage{amsmath}
\usepackage{ulem}
\usepackage{bm}
\usepackage{graphics}
\usepackage[dvips]{graphicx,color}
\usepackage{dcolumn}

\begin{document}

\title{Equilibrium Properties of Trapped Dipolar Gases at Finite Temperatures}
\author{Yuki Endo$^1$, Takahiko Miyakawa$^2$ and Tetsuro Nikuni$^1$}
\affiliation{$^1$Department of Physics, Faculty of Science, Tokyo University of Science, 
1-3 Kagurazaka, Shinjuku-ku, Tokyo 162-8601, Japan\\
$^2$Faculty of Education, Aichi University of Education, 1 Hirosawa, Igaya-cho, Kariya, Aichi 448-8542, Japan}
\date{\today}

\begin{abstract}
We study the equilibrium properties of dipolar Bose and Fermi gases at finite temperatures. We recently developed a variational ansatz for the phase-space distribution function of a dipolar Fermi gas at finite temperatures. We extend the variational formalism to a Bose gas and discuss the effect of dipolar interactions with different statistics on the thermal equilibrium, with particular emphasis on the deformation in momentum space. We examine the stability of the system by varying the temperature, trap aspect ratio and the dipole moment. In addition, we discuss how deformation in both real and momentum space can be observed in the high-temperature regime, which is relevant for the current experiments at JILA [Ni et al. Science {\bf 322}, 231 (2008)] and at University of Tokyo [Aikawa et al. New. J. Phys. {\bf 11}, 055035 (2009)].
\end{abstract}

\maketitle
Interest in dipolar gases has been growing since Bose--Einstein condensates (BECs) of ${\rm ^{52}Cr}$ atoms, which have large magnetic dipole moments, were experimentally observed~\cite{Griesmaier2005,Stuhler2005,Giovanazzi2006}. The anisotropic and long-range nature of the dipolar interaction confers interesting properties to the equilibria and dynamics of dipolar gases. A number of experiments on $^{52}{\rm Cr}$ BECs have investigated their ground states and expansion~\cite{Griesmaier2005,Stuhler2005,Lahaye2007,Koch2008}. On the other hand, several experiments are proceeding energetically toward creating heteronuclear polar molecules whose large electric dipole moments give rise to strong dipolar interactions. An example of a dipolar Fermi molecule is $^{87}{\rm Rb}^{40}{\rm K}$~\cite{Ni2008} and examples of dipolar Bose molecules are $^{87}{\rm Rb}^{41}{\rm K}$~\cite{Aikawa2009}, $^{133}{\rm Cs}^{7}{\rm Li}$~\cite{Deiglmayr2008}, and $^{7}{\rm Li}^{40}{\rm K}$~\cite{Aymar2005}. Rydberg atoms have also been attracting interest due to their large electric dipole moments.

There have also been several theoretical studies of dipolar BECs, which have investigated their ground states~\cite{Santos2000,Yi2000}, collective oscillations~\cite{Yi2001,Goral2002_2}, and their properties in optical lattice potentials~\cite{Goral2002,Danshita2008}. There is also growing interest in dipolar Fermi gases. The ground state~\cite{Goral2001,Miyakawa2008} and stability~\cite{Miyakawa2008,Zhang2009} of dipolar Fermi gases has been theoretically studied. In addition, there have been studies of free expansion~\cite{He2008,Nishimura2009}, collective oscillations~\cite{Sogo2009}, and the possibility of superfluid phases~\cite{Baranov2002,Baranov2004,Bruun2008,Zhao2009}.

We are currently focusing on experiments with $^{52}{\rm Cr}$ BECs. For example, the expansion dynamics of $^{52}{\rm Cr}$ BECs have been observed and it was found that the long-range and anisotropic character of the dipolar interaction leads to an anisotropic deformation of expanding $^{52}{\rm Cr}$ BECs~\cite{Griesmaier2005,Stuhler2005}. These observations indicate that the dipolar interaction tends to elongate dipolar BECs in the dipole moment direction and shrink it radially. The aspect ratio of dipolar atomic gases after free expansion is very sensitive to the trap geometry and the ratio between the dipolar and interatomic interactions. The reason why dipolar BECs elongate along the dipole moment, where two individual dipoles attract each other, can be understood by considering the shape of the dipole--dipole potential~\cite{Stuhler2005}. It was found that the dipole--dipole potential within a cloud of atoms is a parabolic saddle with a negative curvature along the dipole moment. Consequently, the total energy of a trapped dipolar BEC is reduced if the atoms are redistributed from the repulsive to the attractive direction, even at the cost of increasing the external trapping potential energy.

As mentioned above, the shape of dipolar BECs strongly depends on the strength of the effective dipolar interaction. Experimentally, one can adjust the strength of effective dipolar interaction by using a Feshbach resonance, which can control the interatomic interaction. By reducing the interatomic interaction, one can enhance the effective dipolar interaction. The aspect ratio of a dipolar BEC after free expansion has been experimentally observed using this technique.~\cite{Lahaye2007}. In this experiment, the large deformation induced by the dipolar interaction was observed by adjusting the scattering length to approximately zero, where the effective dipolar interaction is largest. On the other hand, many theoretical studies have shown that a dipolar BEC becomes more elongated in the dipole moment direction with increasing strength of the effective dipolar interaction~\cite{Goral2000,Yi2000,Santos2000}. In terms of the stability of dipolar BECs, it was found that the critical scattering length strongly depends on the trap aspect ratio. In a cigar-shaped trap, a repulsive interatomic interaction is required to keep the system stable, whereas in an oblate trap, there is stable region with an attractive interatomic interaction because of the effective repulsive dipolar interaction~\cite{Koch2008}.

On the other hand, there is a growing expectation of creating dipolar Fermi gases. Unlike dipolar BECs, single-component dipolar Fermi gases do not interact via {\it s}-wave collisions and they have Hartree direct and Fock exchange energies of the dipolar interaction in the mean-field description, which reflects the antisymmetric many-body wave function. Miyakawa et al. introduced a variational Wigner function to examine the ground state of the system at zero temperature and they demonstrated that the anisotropic nature of the interaction causes Fermi surface deformation through the Fock exchange energy~\cite{Miyakawa2008}. They found that the real-space distribution, which depends on the trap geometry, is elongated more along the dipole moment than the real-space distribution of system without a dipolar interaction and that the shift is largest in an isotropic trap. In addition, they found that the momentum distribution is always elongated in the dipole moment direction by the Fock exchange energy. Their variational calculations are highly suitable for practical calculations~\cite{Ronen2010,Zhang2009}. Due to the Pauli exclusion principle, fermionic systems have much higher energy scales than BECs. Consequently, a larger dipole moment is required to observe dipolar effects in dipolar Fermi gases than in dipolar BECs. Miyakawa et al.~\cite{Miyakawa2008} have shown that for heteronuclear molecules with a typical electric dipole moment of the order of $1$ Debye, dipolar effects can be easily detected by studying the equilibrium properties of dipolar Fermi gases at zero temperature. On the other hand, Sogo et al. studied the dynamics of free expansion of dipolar Fermi gases~\cite{Sogo2009}. They found that the aspect ratios of dipolar Fermi gases after free expansion are stretched along the dipole moment regardless of the trap geometry. This result contrasts to that of dipolar BECs, which strongly depends on the trap aspect ratio~\cite{Stuhler2005,Giovanazzi2006,Lahaye2007}. Some theoretical studies of dipolar Fermi gases have focused on the zero-temperature properties of these gases and there have been recent theoretical studies on finite-temperature gases~\cite{Zhao2009,Zhang2010,Kestner2010,Yuki2010_3,Baillie2010}.

As described above, a huge dipole moment is required to observe dipolar effects in a Fermi system. However, despite many groups conducting experiments, no groups have succeeded in cooling polar molecules to the quantum-degenerate regime~\cite{Ni2008}. It is thus important to investigate the temperature range in which the dipolar interaction has appreciable effects. It is also important to quantify to what extent the dipolar interaction effect can be observed in the temperature regime of the current experiment. For this reason, we studied the equilibrium properties of dipolar Fermi gases above the Fermi temperature~\cite{Yuki2010_3}. We have previously estimated the electric dipole moments and the trap frequencies required to observe deformations in momentum and real space in the high-temperature regime~\cite{Yuki2010_3}. This result indicates that it is possible to observe deformation in real and momentum space in current experiments~\cite{Ni2008}.

In the present paper, we study the equilibrium properties and stability of normal dipolar Bose gases above the Bose--Einstein transition temperature as well as dipolar Fermi gases. In particular, we focus on the different properties induced by the different statistics of dipolar Fermi and Bose gases because many experimental groups are progressing toward creating dipolar Bose molecules as well as dipolar Fermi molecules (e.g., $^{87}{\rm Rb}^{40}{\rm K}$ molecules~\cite{Ni2008} and $^{87}{\rm Rb}^{41}{\rm K}$). We show that the different properties of Bose and Fermi gases arise from their stabilities and the deformation of their momentum distributions.

\section{Theory}

We consider Bose or Fermi heteronuclear molecules. The dipoles are assumed to be polarized along the ${\it z}$ axis due to an external electric field. In the second quantized form, the Hamiltonian for this system is given by
\begin{eqnarray}
\hat{H}&=&\int d\textbf{r}\hat{\Psi}^\dagger\left(\textbf{r}\right)\left[-\frac{\hbar^2}{2m}\nabla_{\textbf{r}}^2+V_{trap}\left(\textbf{r}\right)\right]\hat{\Psi}\left(\textbf{r}\right)\nonumber \\
	&+&\frac{\left(1+\eta \right)}{2}g\int d\textbf{r}\hat{\Psi}^\dagger\left(\textbf{r}\right)\hat{\Psi}^\dagger\left(\textbf{r}\right)\Psi\left(\textbf{r}\right)\Psi\left(\textbf{r}\right)\nonumber \\
	&+&\frac{1}{2}\int d\textbf{r}\int d\textbf{r}^\prime\hat{\Psi}^\dagger\left(\textbf{r}\right)\hat{\Psi}^\dagger\left(\textbf{r}^\prime\right)V_{dd}\left(\textbf{r}-\textbf{r}^\prime\right)\hat{\Psi}\left(\textbf{r}^\prime\right)\hat{\Psi}\left(\textbf{r}\right),\label{Hamiltonian_fermi}
\end{eqnarray}
where $\hat\Psi\left(\textbf{r}\right)$ is the field operator and the hat indicates a second quantized operator. In the above equation, we defined $\eta\equiv \pm1$, where we use the positive sign for the Bose case and the negative sign for the Fermi case. The first term in Eq. (\ref{Hamiltonian_fermi}) describes the Hamiltonian of a single particle in a harmonic trap:
\begin{eqnarray}
V_{trap}\left(\textbf{r}\right)=\frac{m}{2}\left[ \omega_\rho^2\left(x^2+y^2\right)+\omega_z^2 z^2 \right],\label{trap_potential}
\end{eqnarray}
where ${\it m}$ is the particle mass. The second term describes the contact interaction due to {\it s}-wave scattering, which occurs only for Bose gases. ${\it g}$ is the coefficient of the interatomic interaction. The third term describes the two-body interaction Hamiltonian of the dipolar force:
\begin{eqnarray}
V_{dd}\left(\textbf{r}\right)=\frac{d^2}{r^3}\left(1-3z^2/r^2\right).\label{dipolar_potential}
\end{eqnarray} 
Here, $d$ is the coefficient of the dipolar interaction, and $d^2=p^2/4\pi\epsilon_0$, where $p$ is the magnitude of the electric dipole moment~\cite{Miyakawa2008,Sogo2009} and $\epsilon_0$ is the electric permittivity of vacuum.

To consider the phase-space distribution, we introduce the Wigner distribution function: 
\begin{eqnarray}
{W}\left( \textbf{p},\textbf{r} \right) = \int d \textbf{r}^\prime e^{-i\textbf{p} \cdot \textbf{r}^\prime / \hbar }
          \langle\hat \Psi ^\dagger\left( \textbf{r} - \textbf{r}^\prime /2 \right)\hat \Psi\left( \textbf{r} + \textbf{r}^\prime /2 \right)\rangle,\label{Wigner}
\end{eqnarray}
The density distributions in real and momentum space in terms of ${W}\left( \textbf{p},\textbf{r} \right)$
are given by $n\left(\textbf{p}\right)= \int d\textbf{r}\ W\left(\textbf{p},\textbf{r}\right)$
and $n\left(\textbf{r}\right)=\left(2\pi\hbar\right)^{-3}\int d\textbf{p}\ W\left(\textbf{p},\textbf{r}\right)$, respectively.

We consider the thermal equilibrium of a dipolar gas trapped in a harmonic potential. The system is not globally stable when a dipolar interaction is present because the interaction is partially attractive and causes collapse of the gas. However, a local metastable state exists at $T=0$ and at finite temperatures under certain conditions. As in Ref.~\cite{Yuki2010_3}, we look for this metastable state by introducing a variational Wigner distribution function. We assume the Maxwell--Boltzmann regime at relatively high temperatures, namely
\begin{eqnarray}
W\left(\textbf{p},\textbf{r}\right)=\exp\left\{-\left[\frac{\theta^2}{2m}\left(\frac{p_\rho^2}{\alpha}+\alpha^2p_z^2\right)+\frac{\lambda^2m\omega^2}{2}\left(\beta\rho^2+\frac{z^2}{\beta^2}\right)-\mu_0\right]/k_BT\right\},\label{Wigner2}
\end{eqnarray}
where $\rho^2\equiv x^2+y^2$, $p_\rho^2\equiv p_x^2+p_y^2$, and $\omega\equiv\left(\omega_\rho^2\omega_z\right)^{1/3}$. 
Here, the positive parameters ${\alpha}$ and $\beta$ respectively represent deformations of the density distributions in momentum and real space, and $\lambda$ describes isotropic compression in real space. These parameters are also introduced for $T=0$ in Ref.~\cite{Miyakawa2008}. At finite temperatures, we introduce an additional variational parameter $\theta$ that characterizes isotropic compression in momentum space. The chemical potential is determined by the number constraint $N=\int d\textbf{r}\ n\left(\textbf{r}\right)$, which gives
\begin{eqnarray}
e^{\mu_0/k_BT}=N\left(\frac{\lambda\theta\omega\hbar}{k_BT}\right)^3.\label{mu}
\end{eqnarray}

From the normalization condition
\begin{eqnarray}
N=\int d\textbf{r}\int\frac{d\textbf{p}}{\left(2\pi\hbar\right)^3}W\left(\textbf{p},\textbf{r}\right),
\end{eqnarray}
the chemical potential $\mu_0$ is determined by Eq. (\ref{mu}). Under this ansatz, the density distribution in momentum space is given by
\begin{equation}
n\left(\textbf{p}\right)=N\left(\frac{2\pi\hbar^2\theta^2}{mk_BT}\right)^{3/2}\exp\left[ -\frac{\theta^2}{2mk_BT}\left( \frac{p_\rho^2}{\alpha}+\alpha^2p_z^2 \right)\right].\label{MomentumDistribution}
\end{equation}
Thus, the aspect ratio of the momentum space distribution,
which is the ratio of the root-mean-square momentum in the $p_z$ direction to that in a direction
in the $p_x$--$p_y$ plane (we choose the $p_x$ direction in the following) $\sqrt{\langle p_z^2\rangle/\langle p_x^2\rangle}$,
becomes $\alpha^{-3/2}$.
Similarly, we obtain the density distribution in real space as
\begin{equation}
n\left(\textbf{r}\right)=N\left( \frac{2\pi m\omega^2\lambda^2}{k_BT} \right)^{3/2}\exp\left[ -\frac{m\omega^2\lambda^2}{2k_BT}\left(\beta\rho^2+\frac{z^2}{\beta^2} \right)\right],\label{RealDistribution}
\end{equation}
leading to the aspect ratio of the real-space distribution (i.e., the ratio of the root-mean-square radius in the $z$ direction to that in the $x$ direction) being $\sqrt{\langle z^2\rangle/\langle x^2\rangle}=\beta^{3/2}$.
We note that, from Eq. (\ref{RealDistribution}), $\lambda>1$ ($\lambda < 1$) corresponds to compression (expansion) of the gas in real space, which is a consequence of the effective attraction (repulsion) of the dipolar interaction. In addition, comparing Eq. (\ref{MomentumDistribution}) with Eq. (\ref{RealDistribution}) reveals that $\theta$ plays the same role in real space as $\lambda$ plays in momentum space.

To find a metastable state at a temperature $T$, we look for a local minimum of the Helmholtz free energy
\begin{eqnarray}
F=E-TS.
\end{eqnarray}
Here, $E$ is the total energy, and the entropy $S$ is given by
\begin{eqnarray}
S&=&-k_B\int d\textbf{r}\int \frac{d\textbf{p}}{\left(2\pi\hbar\right)^3}\left\{W\left(\textbf{p},\textbf{r}\right)\ln\left[ W\left(\textbf{p},\textbf{r}\right) \right]\mp\left[1\pm W\left(\textbf{p},\textbf{r}\right)\right]\ln\left[1\pm W\left(\textbf{p},\textbf{r}\right)\right]\right\}\nonumber \\
	&\simeq&-k_B\int d\textbf{r}\int\frac{d\textbf{p}}{\left(2\pi\hbar\right)^3}W\left(\textbf{p},\textbf{r}\right)\ln\left[W\left(\textbf{p},\textbf{r}\right)\right].\label{entropy}
\end{eqnarray}
In the first line of Eq. (\ref{entropy}), the upper and lower signs are for Bose and Fermi gases, respectively. In the above expression for the entropy, we assumed $W\ll 1$, which is consistent with the Maxwell--Boltzmann distribution.

The total energy can be written as a sum of five contributions: the kinetic energy $E_{K}$, the trapping potential energy $E_{V}$, the interatomic mean-field energy $E_g$, which arises only for Bose gases, the Hartree direct energy $E_{H}$, and the Fock exchange energy $E_{ex}$: $E=E_K+E_V+\frac{1+\eta}{2}E_g+E_H+\eta E_{ex}$. The five contributions are given in terms of the Wigner distribution function as~\cite{Yuki2010_3}
\begin{eqnarray}
E_K&=&\frac{1}{2m}\int d\textbf{r}\int \frac{d\textbf{p}}{\left(2\pi\hbar\right)^3}p^2 W\left(\textbf{p},\textbf{r}\right),\label{energy_1}\\
E_V&=&\int d\textbf{r}\int \frac{d\textbf{p}}{\left(2\pi\hbar\right)^3}V\left(\textbf{r}\right) W\left(\textbf{p},\textbf{r}\right),\label{energy_2}\\
E_{g}&=&g\int d\textbf{r}\int\frac{d\textbf{p}}{\left(2\pi\hbar\right)^3}\int \frac{d\textbf{p}^\prime}{\left(2\pi\hbar\right)^3}W\left(\textbf{p},\textbf{r}\right) W\left(\textbf{p}^\prime,\textbf{r}\right),\label{energy_5}\\
E_{H}&=&\frac{1}{2}\int d\textbf{r}\int d\textbf{r}^\prime \int\frac{d\textbf{p}}{\left(2\pi\hbar\right)^3}\int \frac{d\textbf{p}^\prime}{\left(2\pi\hbar\right)^3}V_{dd}\left(\textbf{r}-\textbf{r}^\prime\right) W\left(\textbf{p},\textbf{r}\right) W\left(\textbf{p}^\prime,\textbf{r}^\prime\right),\label{energy_3}\\
E_{ex}&=&\frac{1}{2}\int d\textbf{R}\int d\textbf{s}\int \frac{d\textbf{p}}{\left(2\pi\hbar\right)^3}\int\frac{d\textbf{p}^\prime}{\left(2\pi\hbar\right)^3}V_{dd}\left(\textbf{s}\right)e^{i\left(\textbf{p}-\textbf{p}^\prime\right)\cdot\textbf{s}/\hbar}W\left(\textbf{p},\textbf{R}\right)W\left(\textbf{p}^\prime,\textbf{R}\right).\label{energy_4}
\end{eqnarray}
In Eq. (\ref{energy_4}), we have introduced the center of mass coordinate $\textbf{R}=\left(\textbf{r}+\textbf{r}^\prime\right)/2$ and the relative coordinate $\textbf{s}=\textbf{r}-\textbf{r}^\prime$. In deriving Eqs. (\ref{energy_3}) and (\ref{energy_4}), we used mean-field decoupling:
\begin{eqnarray}
\langle \hat\Psi^\dagger\left(\textbf{r}\right)\hat{\Psi}^\dagger\left(\textbf{r}^\prime\right)\hat{\Psi}\left(\textbf{r}^\prime\right)\hat{\Psi}\left(\textbf{r}\right)\rangle\simeq \langle\hat{\Psi}^\dagger\left(\textbf{r}\right)\hat{\Psi}\left(\textbf{r}\right)\rangle \langle\hat{\Psi}^\dagger\left(\textbf{r}^\prime\right)\hat{\Psi}\left(\textbf{r}^\prime\right)\rangle +\eta \langle\hat{\Psi}^\dagger\left(\textbf{r}\right)\hat{\Psi}\left(\textbf{r}^\prime\right)\rangle\langle\hat{\Psi}^\dagger\left(\textbf{r}^\prime\right)\hat{\Psi}\left(\textbf{r}\right)\rangle,\label{decoupling}
\end{eqnarray}
The difference of the sign of $\eta$ in Eq. (\ref{energy_4}) arises from the different statistics of the two systems. Consequently, the Fock exchange energy has different signs for Bose and Fermi systems.

Inserting the variational function (\ref{Wigner2}) into Eqs. (\ref{energy_1})--(\ref{energy_4}) and (\ref{entropy}), we express the free energy in terms of variational parameters:
\begin{eqnarray}
F&=&\Bigl( \frac{1}{2\theta^2}f\left(\alpha\right)+\frac{1}{2\lambda^2}g\left(\beta\right)+\frac{D\lambda^3}{\left(k_BT\right)^{5/2}}\left\{\frac{1+\eta}{2}g+\frac{\pi d^2}{3}\left[I\left(\beta\right)+\eta I\left(\alpha\right)\right]\right\} \Bigr.\nonumber \\
	& & \ \ \Bigl. 
	-3+\ln\left[N\left(\lambda\theta\omega\hbar\right)^3\right]-3\ln\left[k_BT\right] \Bigr)Nk_BT,\label{FreeEnergy}
\end{eqnarray}
In the above equation, we have defined the scaling functions:
\begin{eqnarray}
f\left(\alpha\right)&=&2\alpha+1/ \alpha^2,\label{scaling_function_1}\\
g\left(\beta\right)&=&2\beta_0/\beta+\beta^2/\beta_0^2,\label{scaling_function_2}
\end{eqnarray}
where $\beta_0\equiv\left(\omega_\rho/\omega_z\right)^{3/2}$. The deformation function $I\left(\alpha\right)$ is defined as 
\begin{eqnarray}
I\left(\alpha\right)\equiv\int_0^\pi d\Theta \sin{\Theta}\left(\frac{3\cos^2\Theta}{\alpha^3\sin^2\Theta+\cos^2\Theta}-1\right).\label{deformation_function}
\end{eqnarray}
The property of $I\left(\alpha\right)$ is described in Appendix~\ref{AppendixA}. In addition, we defined
\begin{eqnarray}
D\equiv \frac{N m^{3/2}\omega^3}{2^3\pi^{3/2}}\label{keisuu_D}
\end{eqnarray}
in Eq. (\ref{FreeEnergy}). The first and second terms in the first line of Eq. (\ref{FreeEnergy}) are the kinetic and potential energies, respectively. The kinetic energy is inversely proportional to the square of $\theta$. Thus, an increase in $\theta$ indicates a reduction in the effective temperature $T^*=T/\theta^2$ associated with the kinetic energy. The third term is the interatomic mean-field energy due to {\it s}-wave scattering. The interatomic mean-field energy contributes only to the compression/expansion in the real-space distribution. Although the mean-field energy is isotropic, it is closely related to the stability of the Bose system, as discussed below. The fourth and fifth terms are respectively the Hartree direct energy and the Fock exchange energy due to the dipolar interaction. Note that the deformation parameters $\alpha$ and $\beta$ contribute independently to the internal energy, $E$. The momentum space deformation parameter $\alpha$ appears only in the kinetic and Fock exchange energies, whereas the real space deformation parameter $\beta$ appears only in the potential and Hartree direct energy. The last three terms in Eq. (\ref{FreeEnergy}) express the entropy, which is independent of $\alpha$ and $\beta$.

\section{Equilibrium Solutions}

We now discuss equilibrium solutions of dipolar quantum gases. The equilibrium solution can be found by minimizing the free energy with respect to the four variational parameters,
$\alpha$, $\beta$, $\lambda$, and $\theta$. To be a local minimum of the free energy representing a metastable state, the solution must be a convex downward point in the four-dimensional space of the variational parameters.
This requires the following conditions:
\begin{eqnarray}
\left\{
\begin{array}{l}
0<\alpha<1,\\
\beta_0<\beta,
\end{array}
\right.\ \left( {\rm Fermi} \right), \ \ \ \ 
\left\{
\begin{array}{l}
1<\alpha,\\
\beta_0<\beta,
\end{array}
\right.\ \left( {\rm Bose} \right).
\label{analyticcondition}
\end{eqnarray}
The above conditions arise from $\partial F/\partial \alpha=0$ and $\partial F/\partial \beta=0$ exploiting the properties of $f\left(\alpha\right)$ and $g\left(\beta\right)$. The difference between Fermi and Bose systems arise from the different signs of the Fock exchange energy.

From the conditions (\ref{analyticcondition}), we find that the momentum space distribution is always elongated in the $p_z$ direction in a Fermi gas, whereas it is stretched in the $p_x$--$p_y$ plane in a Bose gas. In contrast, the real-space distribution tends to be stretched in the $z$ direction and the aspect ratio of the real-space distribution is larger than that for non-dipolar gases in both Fermi and Bose systems.

These different properties can be easily explained. If there is no Fock exchange energy, the conditional equations, $\alpha=1$ and $\theta=1$, apply for both Fermi and Bose systems. The Fock exchange energy leads to deformation and compression of the momentum distribution. Because the sign of the Fock exchange energy differs for Bose and Fermi systems, deformation in momentum space occurs in opposite directions in the two systems.

To find the equilibrium solution, we numerically minimize the free energy by varying three system parameters (temperature $T$, trap aspect ratio $\beta_0$, and the magnitude of the electric dipole moment $p$ or the ratio between the dipolar and interatomic interactions $g/d^2$) for $m=100$ a.u.m., $\omega=2\pi\times 10^2 $ Hz, and $N=10^4$. We normalize the temperature using the ideal gas Fermi temperature $T_F^0=\left(6N\right)^{1/3}\hbar\omega/k_B\simeq188$ nK for a Fermi system and the ideal gas BEC temperature $T_c^0=\left(\hbar\omega N^{1/3}\right)/\left(1.202^{1/3}k_B\right)\simeq 97.2$ nK for a Bose system, which are defined for an isotropic trap.

\subsection{The interaction energy dependencies of the variational parameters}

First, we analyze the variational parameters as functions of the interaction energy.
  \begin{figure}
  \begin{center}
      \scalebox{0.5}[0.5]{\includegraphics{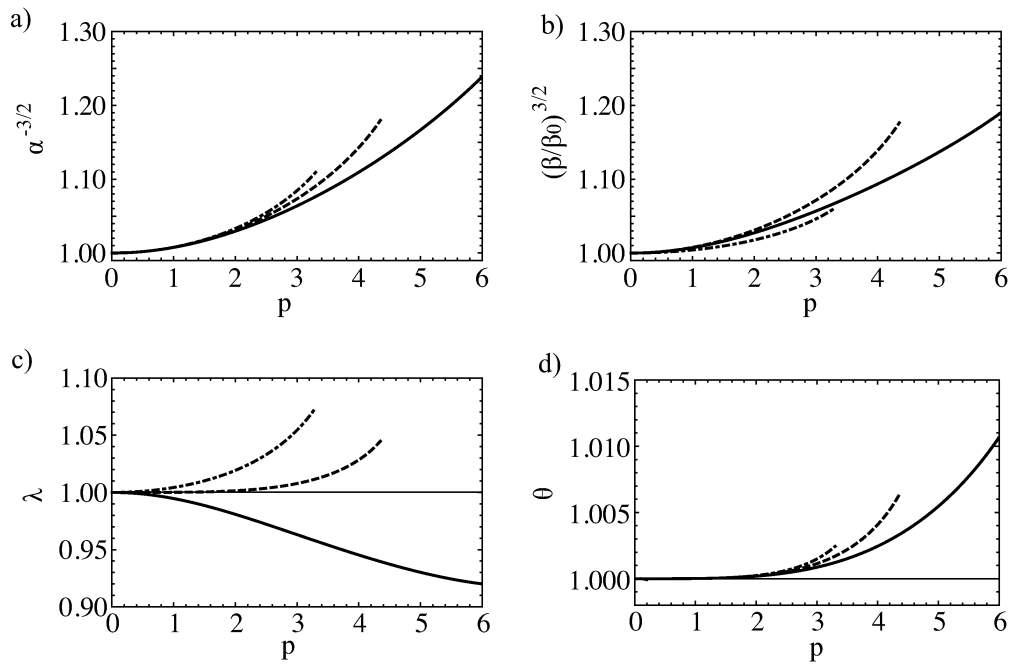}}  
    \caption{Variational parameters of the dipolar Fermi gas as functions of the dipole moment $p$ at temperature ${\it T}{\rm =2.0T_F^0}$ for a trap asymmetry $\beta_0=0.5$ (solid line), $\beta_0=1.0$ (dashed line), and $\beta_0=2.0$ (dash-dotted line).}
    \label{fig:DFb0}
  \end{center}
\end{figure}
Figure~\ref{fig:DFb0} shows plots of the variational parameters of the dipolar Fermi gas as functions of the dipole moment for different trap aspect ratios when the temperature is fixed at ${\it T}{\rm =2}{\it T_F^0}$. For large trap aspect ratios, we plotted the optimized values for $p<p_c$, where $p_c$ is the critical electric dipole moment above which the variational free energy~(\ref{FreeEnergy}) has no local minimum and the system starts to collapse. For example, the critical dipole moment is $p_c=4.36$ Debye for $\beta_0=1.0$ and $p_c=3.30$ Debye for $\beta_0=2.0$. On the other hand, systems with $\beta_0=0.5$ are stable for $p<6.0$ Debye. These results reveal that the system is unstable in highly elongated cigar-shaped traps and that the system tends to become unstable with increasing dipolar interaction. Figure~\ref{fig:DFb0}(a) shows that the momentum distribution is elongated in the dipole direction regardless of the trap aspect ratio. It is important to note that no deformation of the anisotropic momentum distribution is induced without the dipolar interaction. Next, Fig.~\ref{fig:DFb0}(b) shows that with increasing dipolar interaction, the real-space distribution becomes more elongated in the ${\it z}$ direction relative to the real-space distribution without the dipolar interaction. In addition, an isotropic trap gives the largest shift in $\beta$. Next, Fig.~\ref{fig:DFb0}(c) reveals that $\lambda$ changes from $\lambda<1$ to $\lambda>1$ with increasing trap aspect ratio. The reason for this change is that as the trap shape changes from being oblate to being cigar-shaped, the dipolar interaction changes from being repulsive to attractive due to anisotropy of the dipolar interaction.
Finally, from Fig.~\ref{fig:DFb0}(d), we find that the effective temperature decreases as the dipolar interaction increases regardless of the trap aspect ratio because only the kinetic energy and entropy depend on $\theta$ (Eq. (\ref{FreeEnergy})).

\begin{figure}[h]
  \begin{center}
      \scalebox{0.5}[0.5]{\includegraphics{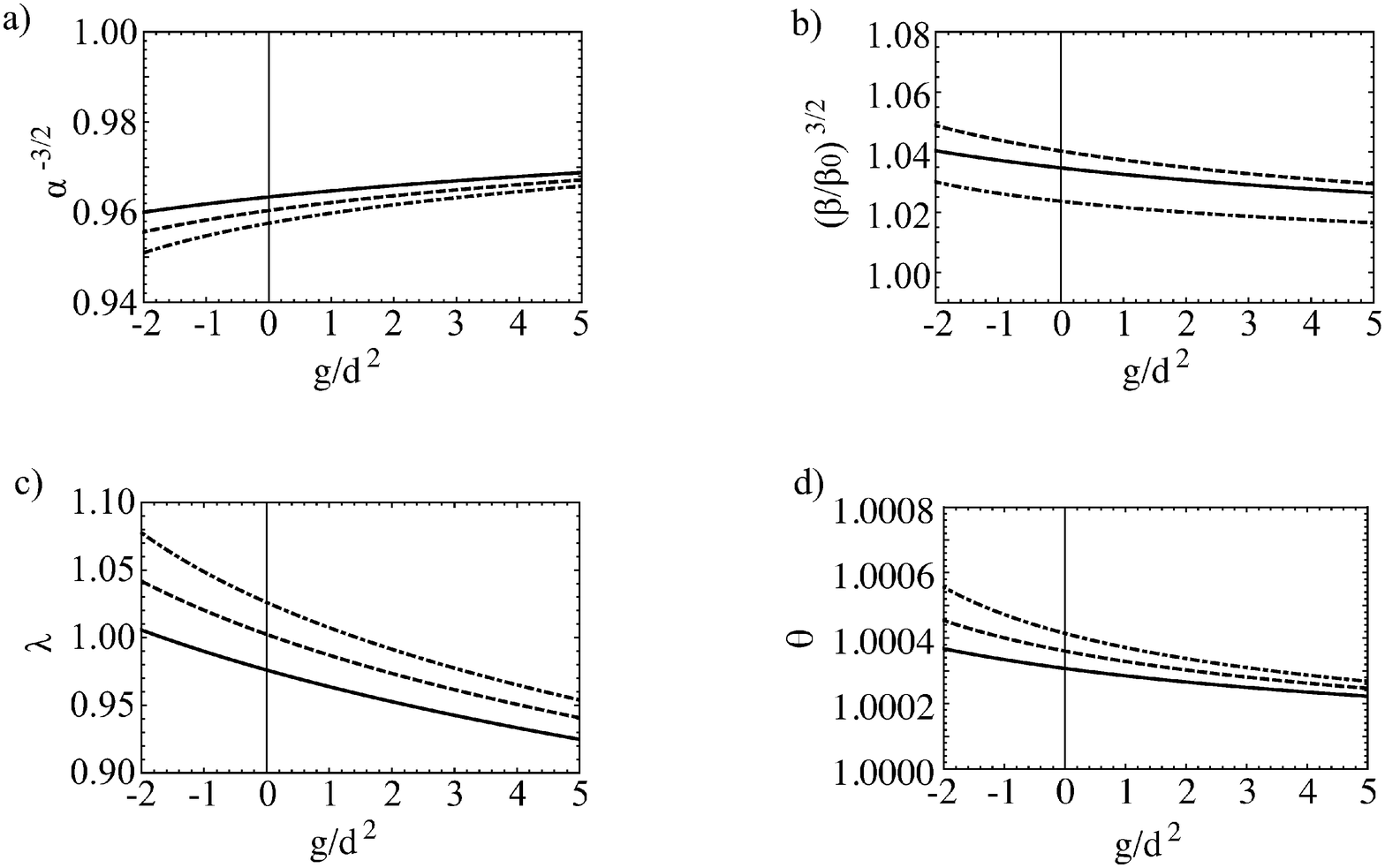}}  
    \caption{Variational parameters of the dipolar Bose gas as functions of the ratio of the interatomic interaction to the dipolar interaction $g/d^2$ at the temperature ${\it T}{\rm =2.0T_c^0}$ for a trap asymmetry $\beta_0=0.5$ (solid line), $\beta_0=1.0$ (dashed line), and $\beta_0=2.0$ (dash-dotted line).}
    \label{fig:DBb0}
  \end{center}
\end{figure}
Figure~\ref{fig:DBb0} shows plots of the scaling parameters of the dipolar Bose gas as functions of the ratio of the interatomic to dipolar interactions $g/d^2$ for different trap aspect ratios with the temperature fixed to ${\it T}{\rm =2}{\it T_c^0}$. Systems with $\beta_0=0.5$, $\beta=1.0$, and $\beta=2.0$ are stable in the region $-2<\left(g/d^2\right)<6$. Figure~\ref{fig:DBb0}(a) shows that the momentum distribution is stretched perpendicular to the dipolar direction contrary to the case of the dipolar Fermi gases. In contrast, Fig.~\ref{fig:DBb0}(b) reveals that with increasing effective dipolar interaction, the real-space distribution becomes more elongated in the $z$ direction relative to the case without the dipolar interaction and that $\beta$ is shifted the most in an isotropic trap, just as for dipolar Fermi gases. Next, Fig.~\ref{fig:DBb0}(c) shows that with decreasing interaction ratio, $\lambda$ shifts to $\lambda>1$. This is because the large positive interaction ratio indicates that the interatomic interaction is large and repulsive, which expands the distribution ($\lambda<1$). On the other hand, the large negative interaction ratio means that the interatomic interaction is large and attractive, which compresses the distribution ($\lambda>1$). In addition, the density increases with decreasing interaction ratio, which enhances the effective dipolar interaction. Finally, from Fig.~\ref{fig:DBb0}(d), we find that the effective temperature decreases as the dipolar interaction increases regardless of the trap aspect ratio because only the kinetic energy and entropy depend on $\theta$ (Eq. (\ref{FreeEnergy})).

Comparing Figs.~\ref{fig:DFb0} and \ref{fig:DBb0}, we find that the different statistics of fermions and bosons manifests itself only as deformation in momentum space and compression in real space. First, dipolar Fermi and Bose gases have opposite properties in momentum space because they have opposite signs for the Fock exchange energy. Next, because only Bose systems have an interatomic interaction, which contributes to real-space compression, dipolar Bose gases depend on the sign of the ratio of the interatomic interaction to the dipolar interaction in addition to their strengths. The interatomic interaction has a significant effect on the stability, which we show later.

\subsection{The temperature dependence of the variational parameters}

\subsubsection{The temperature dependence of the variational parameters for several trap aspect ratios}

Next, we analyze the variational parameters as a function of temperature.
  \begin{figure}
  \begin{center}
      \scalebox{0.5}[0.5]{\includegraphics{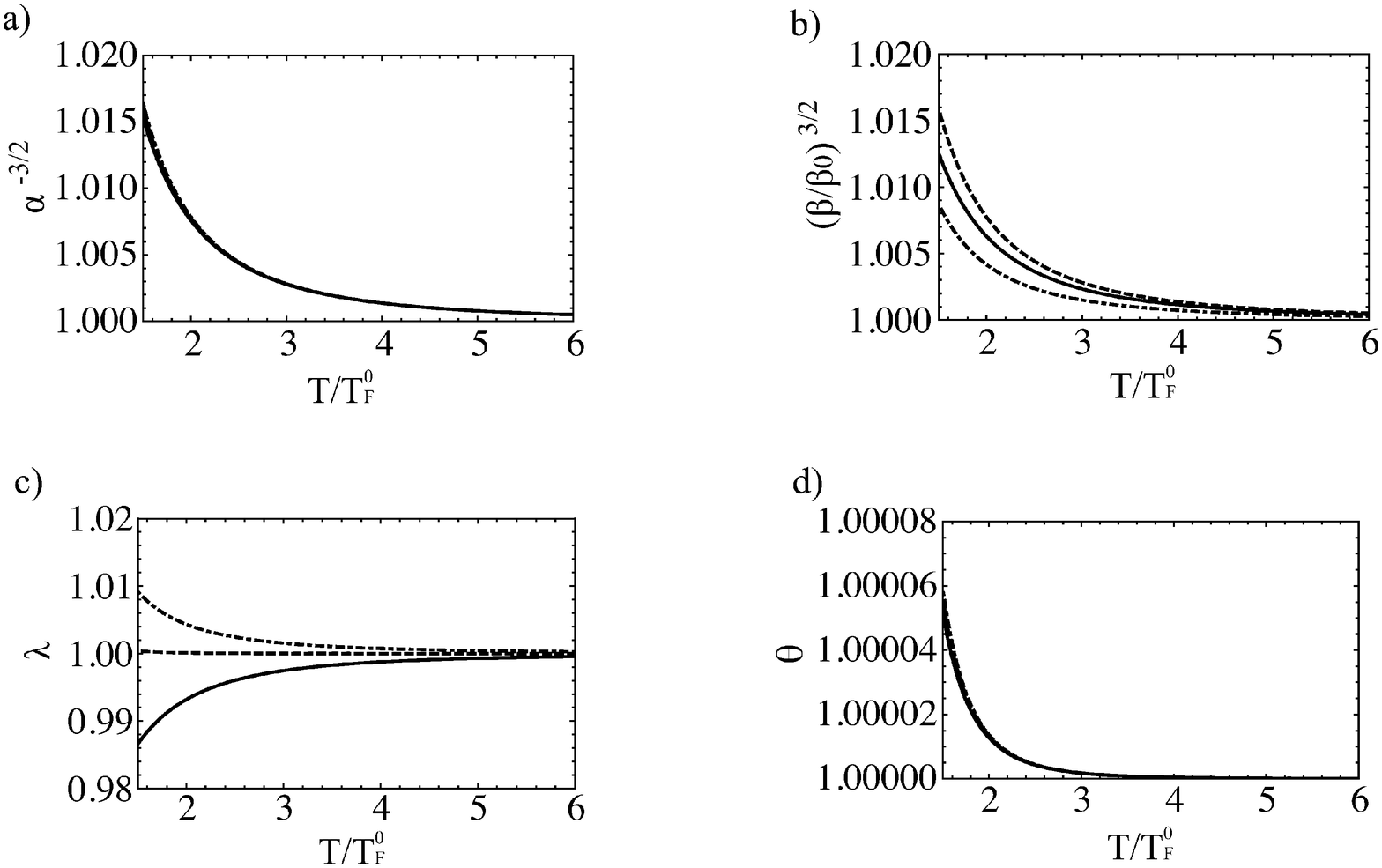}}
       \caption{Variational parameters of the dipolar Fermi gas as functions of the temperature $T$ for a fixed dipole moment $p=1 {\rm \ Debye}$ and for trap asymmetries of $\beta_0=0.4$ (solid line), $\beta_0=1.0$ (dashed line), and $\beta_0=2.0$ (dash-dotted line).}
    \label{fig:DFTb0}
  \end{center}
\end{figure}
Figure~\ref{fig:DFTb0} shows a plot of the variational parameters of the dipolar Fermi gas as functions of temperature for $\beta_0=0.4$ (solid line), $\beta_0=1.0$ (dashed line), and $\beta_0=2.0$ (dash-dotted line), where the electric dipole moment is fixed at $p=1 {\rm \ Debye}$. Figure~\ref{fig:DFTb0}(a) indicates that the momentum distribution becomes more elongated in the dipole direction as the temperature decreases. The deformation in the momentum distribution arises from competition between the kinetic energy, which favors an isotropic momentum distribution, and the exchange energy, which favors an anisotropic momentum distribution. With decreasing temperature, the kinetic energy decreases and the interaction energy increases, increasing the aspect ratio in the momentum distribution.
Figure~\ref{fig:DFTb0}(b) reveals that a lower temperature increases the deviation of the aspect ratio in real-space from that of the non-interacting case. It also shows that the isotropic trap has the largest shift of $\beta$.
Figure~\ref{fig:DFTb0}(c) shows that the magnitude of the deviation from $1$ in $\lambda$ increases at lower temperatures. In addition, we have $\lambda>1$ for $\beta_0>1$ (cigar-shaped trap) and $\lambda<1$ for $\beta_0<1$ (oblate trap).
The Hartree direct energy becomes attractive (repulsive) in the former (latter) case due to deformation of the spatial density distribution, which is mainly determined by the trap geometry. This is leads to compression (expansion) of the gas cloud for $\beta_0>1$ ($\beta_0<1$).
Finally, Fig.~\ref{fig:DFTb0}(d) reveals that the dipolar Fermi gas is compressed in momentum space irrespective of the trap aspect ratio. The interaction effect increases with decreasing temperature in all cases, whereas the deformations in real and momentum spaces are negligibly small for the case $T>2T_F^0$, $p=1\ {\rm Debye}$, and $\omega=2\pi\times 10^2$ Hz.

\begin{figure}[h]
  \begin{center}
      \scalebox{0.5}[0.5]{\includegraphics{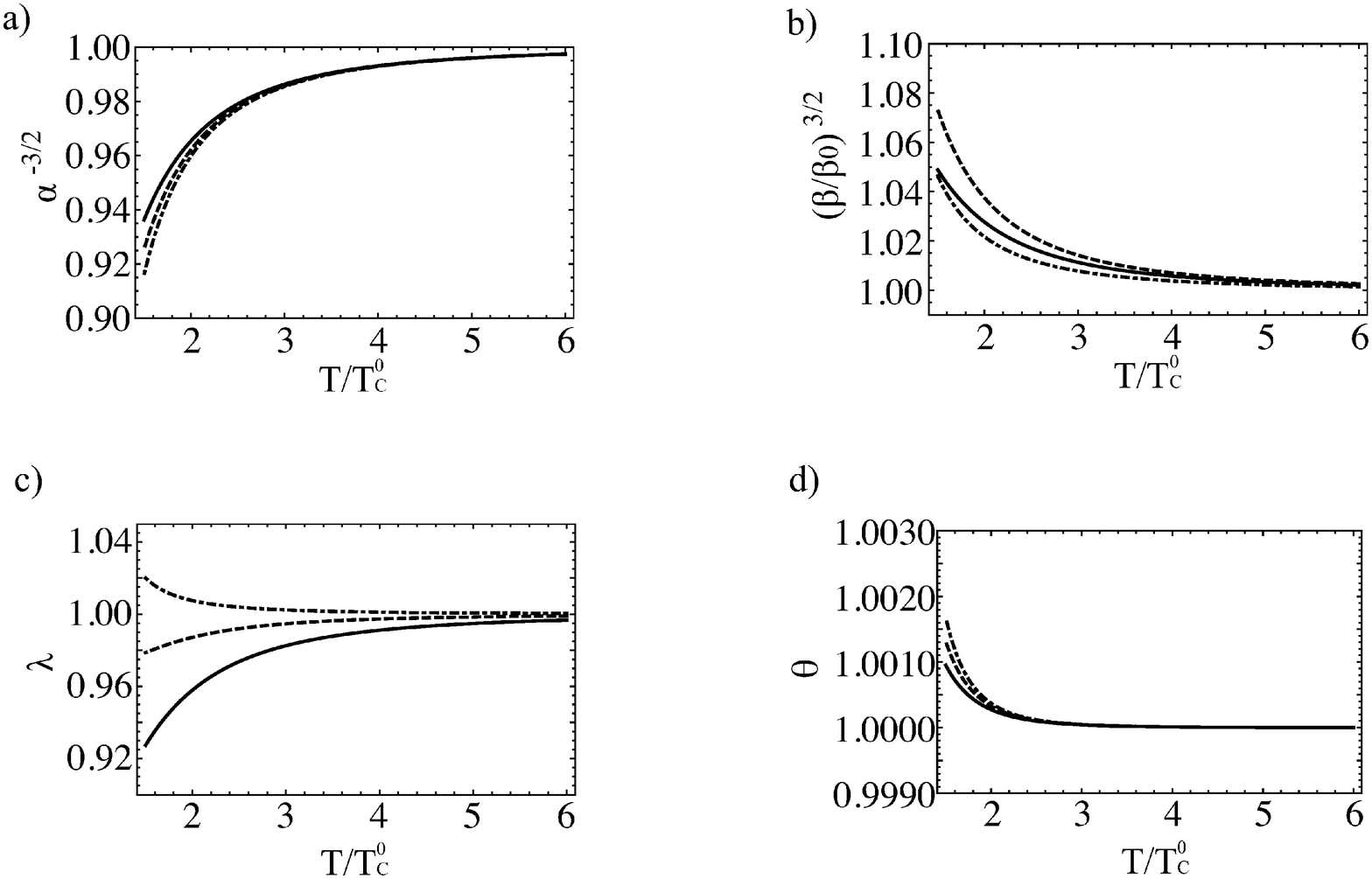}}  
    \caption{Variational parameters of the dipolar Bose gas as functions of the temperature $T$ for a fixed ratio between the dipolar and interatomic interactions $g/d^2=1.0$ for trap asymmetries of $\beta_0=0.4$ (solid line), $\beta_0=1.0$ (dashed line), and $\beta_0=2.0$ (dash-dotted line).}
    \label{fig:DBTb0}
  \end{center}
\end{figure}
Figure~\ref{fig:DBTb0} shows a plot of the variational parameters of the dipolar Bose gas as functions of temperature for $\beta_0=0.4$ (solid line), $\beta_0=1.0$ (dashed line), and $\beta_0=2.0$ (dash-dotted line) and for when the interaction ratio is fixed at $\left(g/d^2\right)=1.0$. Figure~\ref{fig:DBTb0}(a) indicates that the momentum distribution becomes more stretched perpendicular to the dipole direction as the temperature decreases, which is the contrary to the Fermi case. Figure~\ref{fig:DBTb0}(b) reveals that the deformation in real space is the same as that for a dipolar Fermi gas. Figure~\ref{fig:DBTb0}(c) shows that the magnitude of the deviation in $\lambda$ increases at lower temperatures because the effective dipolar interaction increases with decreasing temperature. As the dipolar interaction depends on the trap aspect ratio, we have $\lambda>1$ for $\beta_0>1$ and $\lambda<1$ for $\beta_0<1$. Finally, Fig.~\ref{fig:DBTb0}(d) reveals that a dipolar Fermi gas is compressed in momentum space irrespective of the trap aspect ratio.

\subsubsection{The temperature dependence of the variational parameters for different interaction strengths}

 \begin{figure}
  \begin{center}
      \scalebox{0.5}[0.5]{\includegraphics{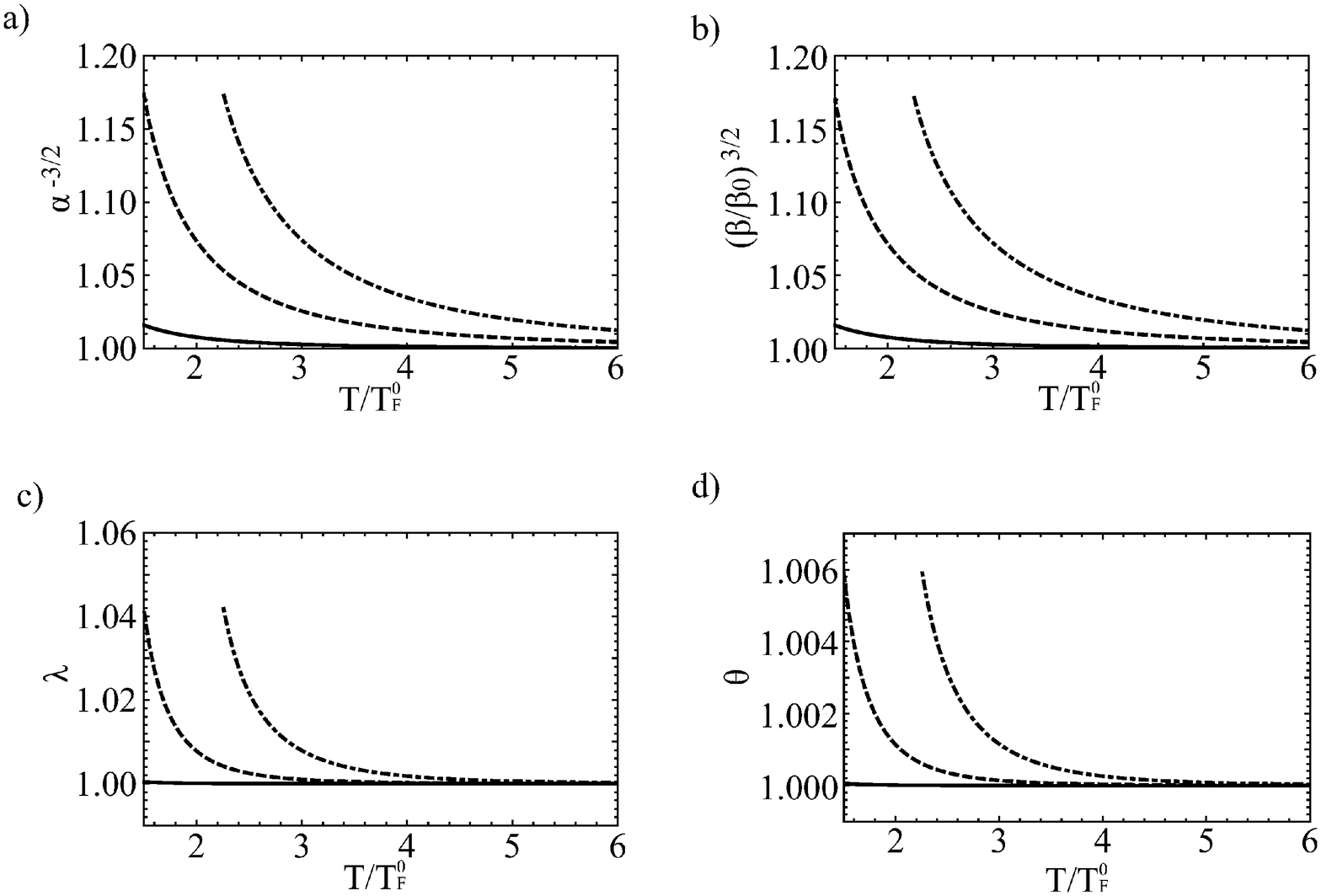}} 
      \caption{Variational parameters of the dipolar Fermi gas as functions of the temperature for a fixed trap asymmetry of $\beta_0=1.0$ for a dipole moment $p=1.0 {\rm \ Debye}$ (solid line), $p=3.0 {\rm \ Debye}$ (dashed line), and $p=5.0 {\rm \ Debye}$ (dash-dotted line).}
    \label{fig:DFTp}
  \end{center}
\end{figure}
Figure~\ref{fig:DFTp} shows the variational parameters of the dipolar Fermi gas as functions of temperature for $p=1.0$ Debye (solid line), $p=3.0$ Debye (dashed line), and $p=5.0$ Debye (dash-dotted line) for an isotropic trap $\beta_0=1.0$. In all cases, the deviation of the variational parameters from those of the non-interacting case becomes more pronounced at lower temperatures and at higher electric dipole moments.
For large electric dipole moments, we plotted the optimized values for $T>T_c$. The critical temperature for this collapse for $p=5.0$ Debye is $T_c={2.25}T_F^0$. On the other hand, systems with $p=1.0$ Debye and $p=3.0$ Debye are stable in the temperature regime $T>1.5T_F^0$. These results reveal that the system is unstable even in the high-temperature regime for large electric dipole moments. Figures~\ref{fig:DFTp}(a) and (b) show that the aspect ratios in both momentum and real spaces are sufficiently large to observe deformation effects in experiments for polar molecules with large electric dipole moments.

\begin{figure}[h]
  \begin{center}
      \scalebox{0.4}[0.4]{\includegraphics{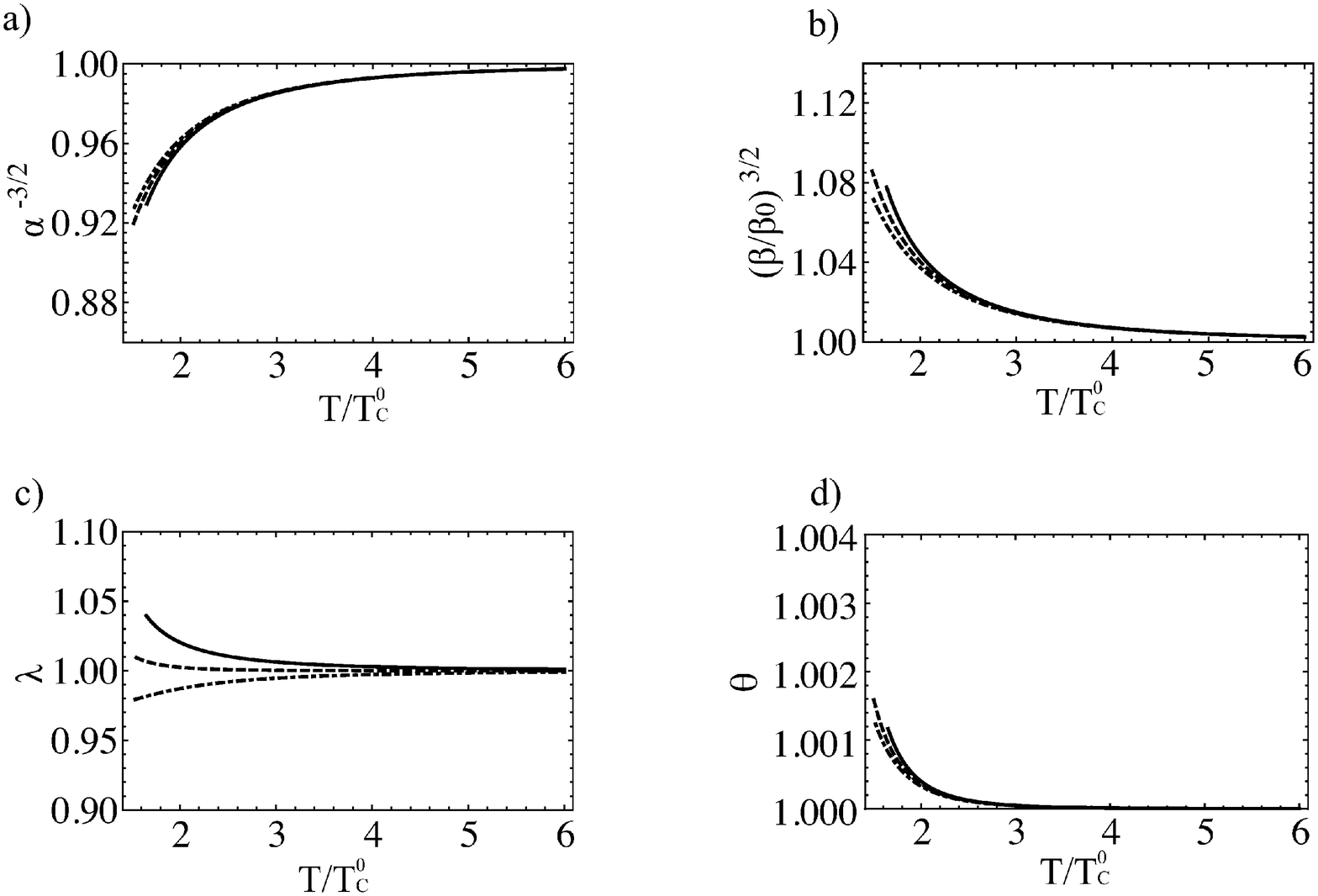}}  
    \caption{Variational parameters of the dipolar Bose gas as functions of the temperature for a fixed trap asymmetry $\beta_0=1.0$ for a ratio between the dipolar and interatomic interactions $g/d^2=-1.0$ (solid line), $g/d^2=0.0$ (dashed line), and $g/d^2=1.0$ (dash-dotted line).}
    \label{fig:DBTUU}
  \end{center}
\end{figure}
Figure~\ref{fig:DBTUU} shows the variational parameters of the dipolar Bose gas as a function of temperature for $g/d^2=-1.0$ (solid line), $g/d^2=0.0$ (dashed line), and $g/d^2=1.0$ (dash-dotted line) for an isotropic trap $\beta_0=1.0$. In all cases, the deviation of the variational parameters from those for the non-interacting case becomes more pronounced at lower temperatures and at smaller ratios of the interactions. This is because a small interaction ratio indicates a small repulsive or attractive interatomic interaction, which shortens the atomic distance and enhances the dipolar interaction. We plotted the optimized values for $T>T_c$ for a small interaction ratio. The critical temperature for this collapse is $T_c={1.62}T_c^0$ for $g/d^2=-1.0$. On the other hand, $g/d^2=0.0$ and $g/d^2=1.0$ are stable in the temperature regime $T>1.5T_c^0$. In addition, from Fig.~\ref{fig:DBTUU}, we see that the magnitude of the deviation from $1$ in $\lambda$ depends on the sign of the interaction ratio at low temperatures.

\section{Stability}

\subsection{Fermi gas}

In this section, we discuss the stability of dipolar gases. We first consider a Fermi gas.
 \begin{figure}
  \begin{center}
      \scalebox{0.5}[0.5]{\includegraphics{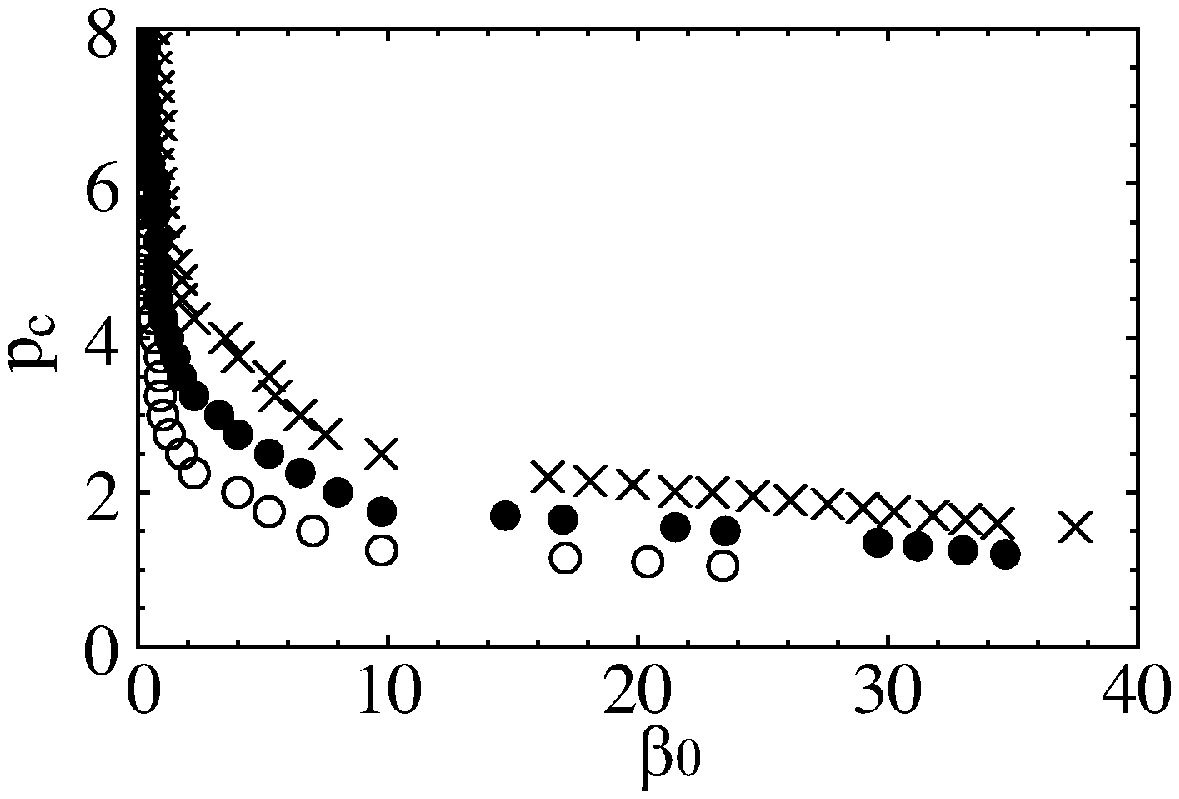}} 
    \caption{Critical value of the dipole moment $p_c$ of the dipolar Fermi gas as a function of the trap aspect ratio for
    $T/T_F^0=1.5$ (open circles), $T/T_F^0=2.0$ (filled circles), and $T/T_F^0=3.0$ (crosses).}
  \label{fig:DF_instability_1}
  \end{center}
  \end{figure}
\begin{figure}
  \begin{center}
      \scalebox{0.5}[0.5]{\includegraphics{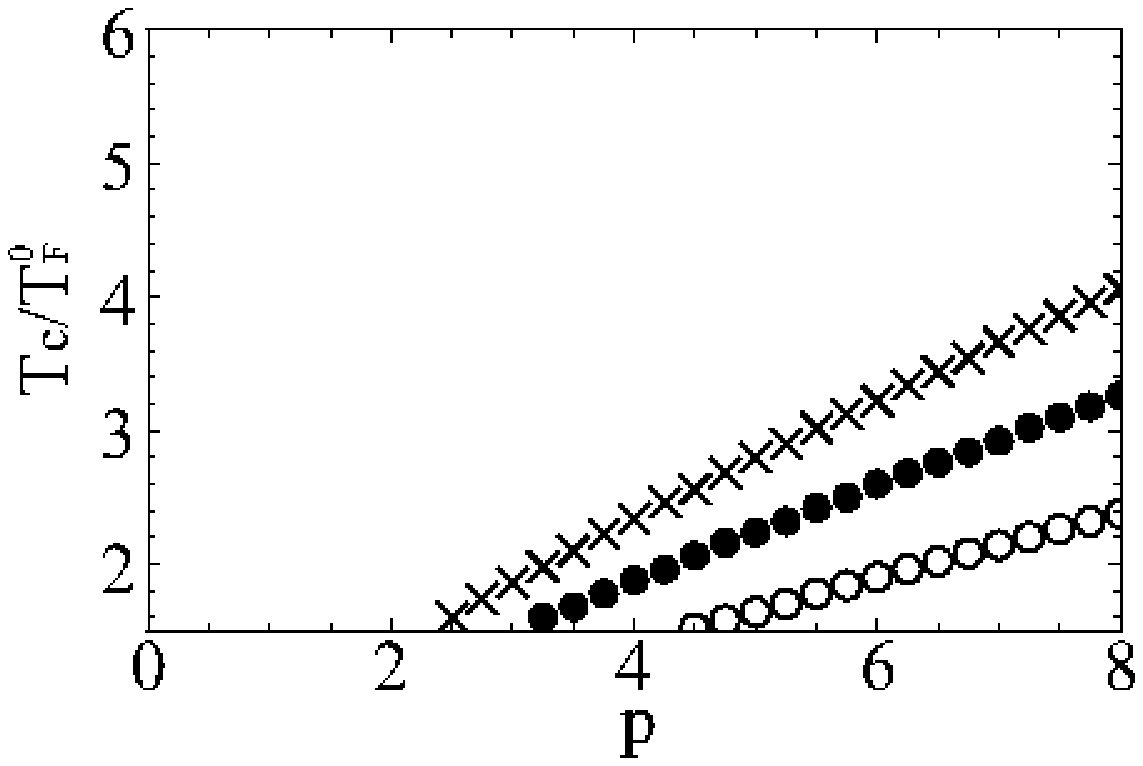}} 
      \scalebox{0.5}[0.5]{\includegraphics{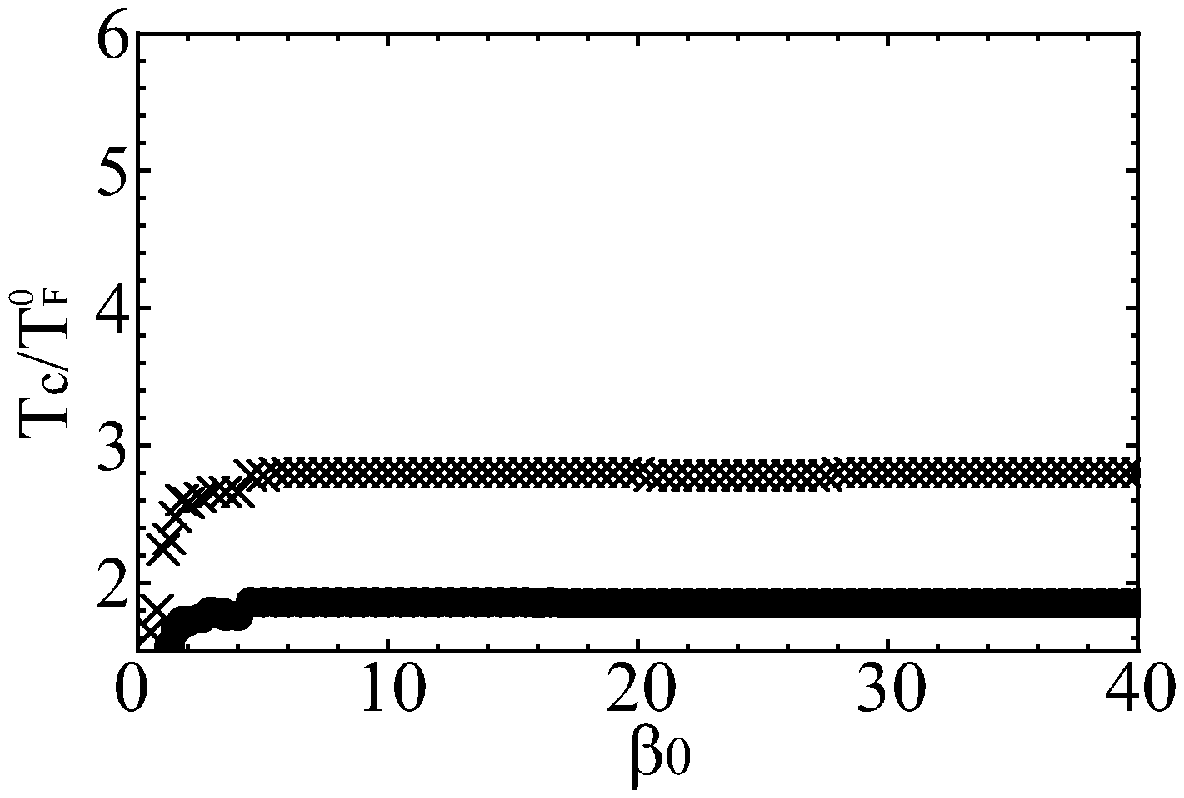}} 
    \caption{Critical temperature of the dipolar Fermi gas. (a) Critical temperature as a function of the dipole moment for trap aspect ratios of $\beta_0={\rm 0.5}$ (open circles), $\beta_0={\rm 1.0}$ (filled circles), and $\beta_0={\rm 2.0}$ (crosses). (b) Critical temperature as a function of the trap aspect ratio for dipole moments of $p={\rm 3.0}$ (filled circles) and $p={\rm 5.0}$ (crosses). 
    This small dip between $\beta_0=20$ and $28$ for $p={\rm 5.0}$ is due to numerical uncertainty.}
  \label{fig:DF_instability_2_3}
  \end{center} 
  \end{figure}
Figure~\ref{fig:DF_instability_1} shows the stability diagram in which the critical values of the electric dipole moment  $p_c$ are plotted as functions of the trap aspect ratio $\beta_0$ for different temperatures.
This result indicates that the critical dipole moment $p_c$ increases drastically as the trap becomes more oblate. This is consistent with the result for a zero-temperature Fermi gas~\cite{Miyakawa2008}. The unstable region expands with decreasing temperature. 
Based on Fig.~\ref{fig:DFTp}, we assume that the critical temperature for collapse instability increases as the electric dipole moment increases. This behavior is confirmed by Fig.~\ref{fig:DF_instability_2_3}(a), which shows the critical temperature $T_c$ for $\beta_0=0.5$, $\beta_0=1.0$, and $\beta_0=2.0$.
Figure~\ref{fig:DF_instability_2_3}(b) shows the critical temperature as a function of the trap aspect ratio for $p=3.0$ and $p=5.0$ Debye. This result indicates that the critical temperature $T_c$ increases as $\beta_0$ increases.
When ${\it p}{\rm =1.0 \ Debye}$, the system is always stable within the parameter range shown in the figure ($\beta_0 \leq 40$). Figure~\ref{fig:DF_instability_2_3}(b) suggests that we should be careful not to enter the unstable region when the temperature is reduced. Even if the system is initially in a stable region at a high temperature, the system may (depending on the trap aspect ratio) become unstable on cooling before reaching the quantum degenerate regime. In addition, the local minimum always disappears when $\lambda$ exceeds a critical value. This suggests that the instability first occurs in real space and a dipolar Fermi gas collapses when $\lambda$ reaches the critical value.

\subsection{Bose gas}

 \begin{figure}
  \begin{center}
      \scalebox{0.5}[0.5]{\includegraphics{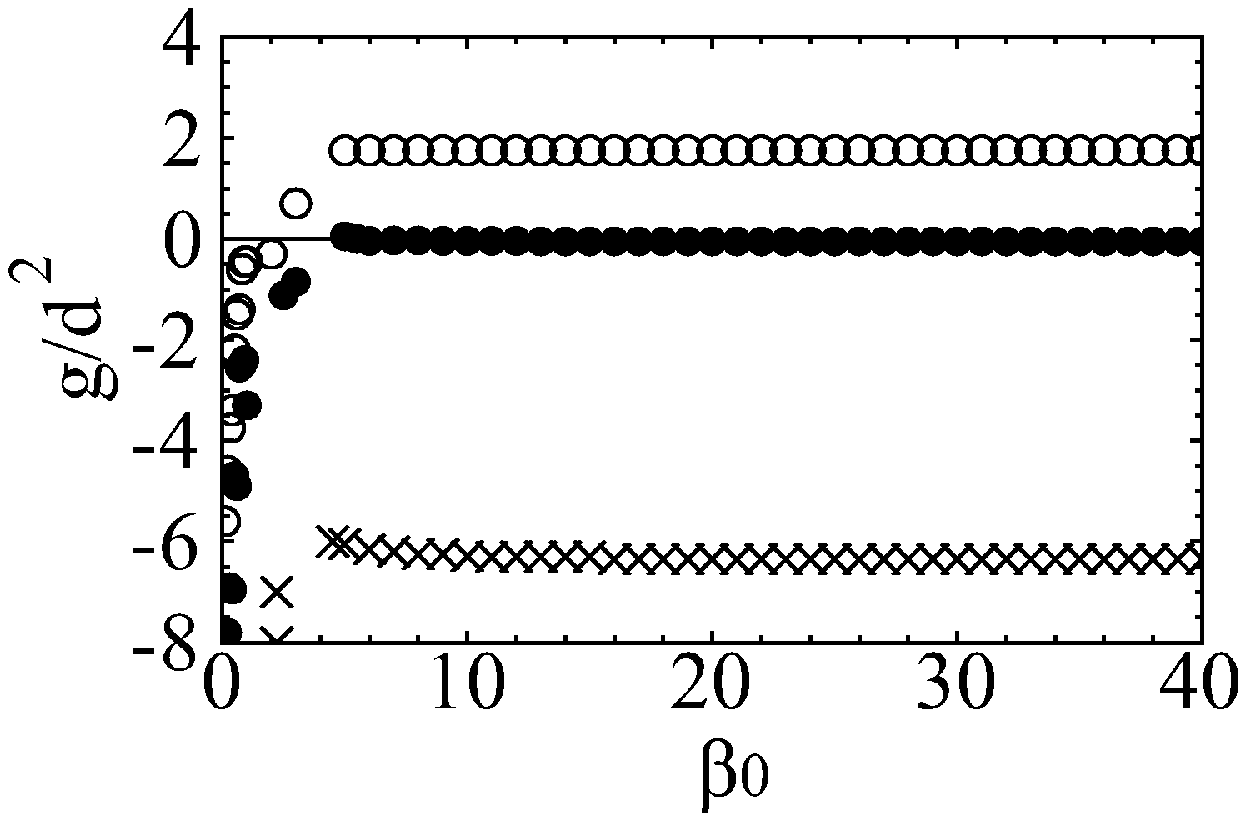}} 
    \caption{Critical value of the ratio between the dipolar interaction and the mean-field energy $\left(g/d^2\right)_c$ of the dipolar Bose gas as a function of the trap aspect ratio for
    $T/T_c^0=1.5$ (open circles), $T/T_c^0=2.0$ (filled circles), and $T/T_c^0=3.0$ (crosses).}
  \label{fig:DB_instability_1}
  \end{center}
  \end{figure}
\begin{figure}
  \begin{center}
      \scalebox{0.5}[0.5]{\includegraphics{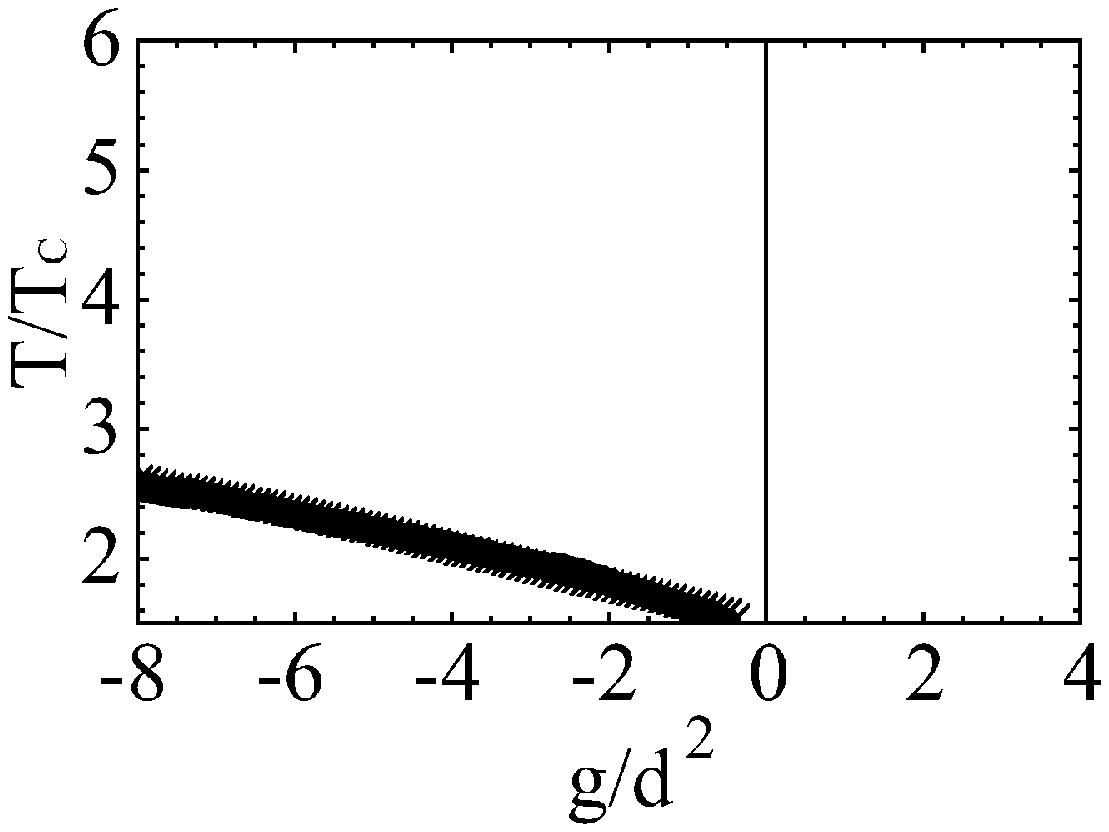}} 
      \scalebox{0.5}[0.5]{\includegraphics{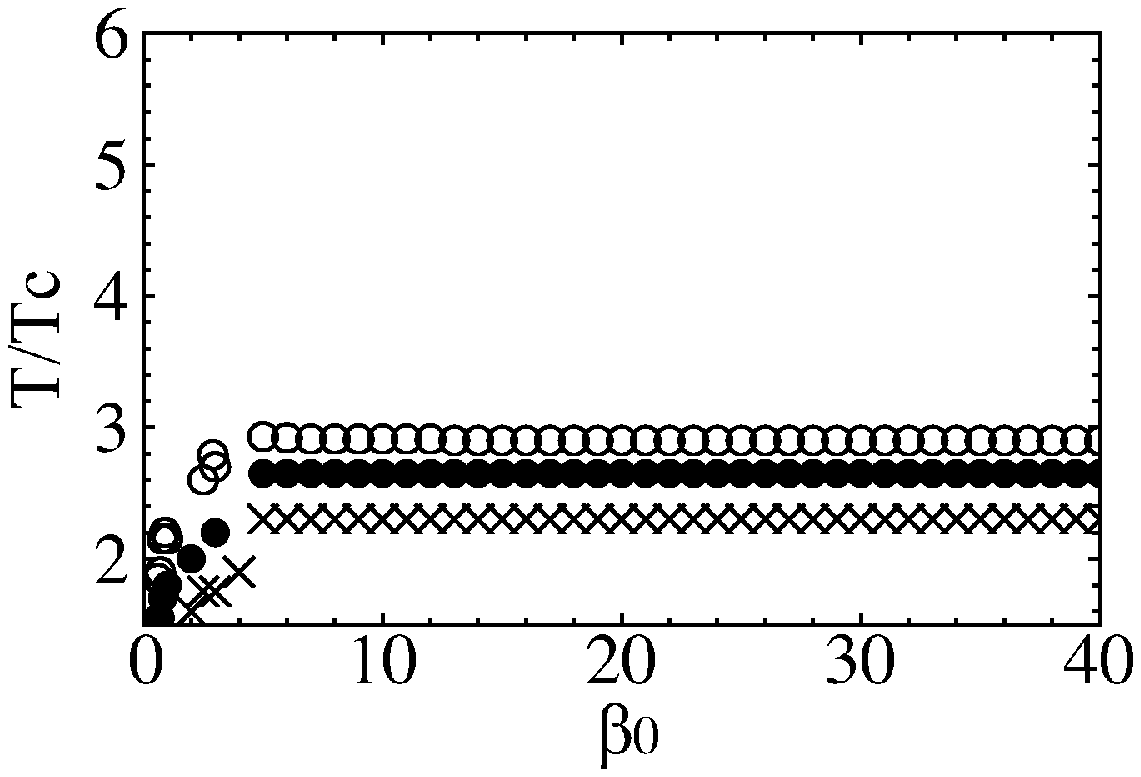}} 
    \caption{Critical temperature of the dipolar Bose gas. (a) Critical temperature as a function of the ratio between the dipolar interaction and the mean-field energy for trap aspect ratios $\beta_0={\rm 1.0}$ (filled circles) and $\beta_0={\rm 2.0}$ (crosses). (b) Critical temperature as a function of the trap aspect ratio for the ratio between the dipolar interaction and the mean-field energy $g/d^2={\rm -1.0}$ (open circle), $g/d^2={\rm 0.0}$ (filled circles), and $g/d^2={\rm 1.0}$ (crosses).}
  \label{fig:DB_instability_2_3}
  \end{center} 
  \end{figure}
Next, we consider Bose gases. Figure~\ref{fig:DB_instability_1} shows the stability diagram in which the critical values of the ratio of the interatomic interaction to the dipolar interaction $\left(g/d^2\right)_c$ are plotted as functions of the trap aspect ratio $\beta_0$ for different temperatures. As the interaction ratio approaches $0$, the dipolar interaction is enhanced, so that the stability becomes more dependent on the trap geometry. Increasing the attractive interatomic interaction causes the system to become unstable. Meanwhile, the effective repulsive dipolar interaction in the oblate trap stabilizes the system. This means that the dipolar interaction keeps the system stable against the attractive interatomic interaction at low trap aspect ratios ($\beta_0<1$). This result is consistent with Fig.~\ref{fig:DBb0}. In Fig.~\ref{fig:DB_instability_2_3}(a), we plot the critical temperatures as functions of the ratio $g/d^2$ for $\beta_0=1.0$ and $\beta_0=2.0$. The dependence of the trap aspect ratio on the stability decreases with decreasing interaction ratio because the instability is caused by the attractive interatomic interaction, and not by the dipolar interaction. When $\beta_0=0.5$, the system is always stable within the parameter range shown in the figure ($-8\leq g/d^2\leq 4$). Figure~\ref{fig:DB_instability_2_3}(b) shows the critical temperature as functions of the trap aspect ratio for different ratios of the interactions. This result indicates that the critical temperature increases with increasing trap aspect ratio. In Fig.~\ref{fig:DB_instability_2_3}(b), the critical temperatures only shift down with increasing ratio of the interaction because expansion or compression of the distribution arising from the interatomic interaction is independent of the trap aspect ratio. In addition, we see that the local minimum always disappears when $\lambda$ exceeds a critical value, as in the case of Fermi gases. This suggests that the instability first occurs in real space for both Fermi and Bose gases.

\section{Relation to Experiments}

We discuss the stability properties in connection with experiments. First, we consider the Fermi case. The JILA experiment~\cite{Ni2008} revealed that the mass of the polar molecule is about ${\it m}{\rm=128 \ a.m.u.}$ and that the electric dipole moment is about ${\it p}{\rm=0.566 \ Debye}$. Taking these values along with $\omega=2\pi\times 10^2$ and $N=10^4$, we find that it is always possible to find a stable dipolar Fermi gas for any temperature $T\geq 1.5 T_F^0$ in the trap aspect ratio regime $\beta_0\leq 40$. Since, according to the results in Ref.~\cite{Miyakawa2008}, a gas of polar molecules at zero temperature with the same system parameters becomes unstable, the critical temperature should lie in $0< T <1.5 T_F^0$. It might be expected that the deformation effect of the distribution can be observed using a highly elongated cigar-shaped trap potential. However, the dipolar Fermi gas becomes unstable at extremely large trap aspect ratios. For $\omega=2\pi\times10^2$, $N=10^4$, $p=0.566$ Debye, and $m=128$ {a.m.u.}, the critical trap aspect ratios for $T=1.5T_F^0$ and $T=2.0T_F^0$ are $\beta_0^c=62.0$ and $\beta_0^c=72.5$, respectively. These effects suggest that deformation effects cannot be observed by just adjusting the trap aspect ratio.

We now discuss how the deformation effects of momentum-space and real-space distributions can be observed in the polar molecules used in the JILA experiment~\cite{Ni2008}.
We assume that $N$, $p$, and $m$ have the same values as those given in the previous paragraph. Increasing $\omega$ causes dipolar interaction effects to become more pronounced because the density of the gas trapped in a tight potential increases and the interaction becomes stronger. For $\omega=5\pi \times 10^2$ Hz, $\beta_0{\rm =1.0}$, and $T=1.5T_F^0$, we have aspect ratios $\sqrt{\langle p_z^2\rangle/\langle p_x^2\rangle}=\alpha^{-3/2}\simeq 1.130$ in the momentum distribution and $\sqrt{\langle z^2\rangle/\langle x^2\rangle}=\beta^{3/2}\simeq 1.127$ in the spatial distribution. For $\omega=6\pi \times 10^2$ Hz, $\beta_0{\rm =1.0}$ and $T=2.0T_F^0$, we have $\alpha^{-3/2}\simeq 1.106$ and $\beta^{3/2}\simeq 1.103$. Thus, the deformation effects arising from the anisotropic nature of the dipolar interaction are detectable using the present experimental conditions.

Next, we consider the Bose case. We assume $^{41}{\rm K}^{87}{\rm Rb}$ molecule~\cite{Aikawa2009}, so we cite the physical quantities from the JILA experiment~\cite{Ni2008}, which uses isomeric molecules: ${\it m}{\rm=128 \ a.m.u.}$ and ${\it p}{\rm=0.566 \ Debye}$. Taking these values along with a scattering length of $a=10$ nm, a coefficient of the dipolar interaction of $d^2\simeq 3.20\times 10^{-50}$, and an interatomic interaction of $g\simeq 6.58\times 10^{-51}$, the ratio of those interactions is $g/d^2\simeq 0.21$. Let us now consider the possibility of experimentally observing the deformation caused by the dipolar interaction. First, we consider a highly elongated trap potential with $\omega=2\pi\times 10^2$ and $N=10^4$. At the temperature $T=2.0T_c^0$, the system is stable for $\beta_0=100.0$ but the deformation in momentum and real spaces scaled by non-dipolar gases are only $\alpha^{-3/2}\simeq 0.981$ and $\left(\beta/\beta_0\right)^{3/2}\simeq 1.000$. On the other hand, at a temperature of $T=1.5T_c^0$, the aspect ratios are $\alpha^{-3/2}\simeq 0.959$ and $\left(\beta/\beta_0\right)^{3/2}\simeq 1.000$ for this trap aspect ratio. These results indicate that using a highly elongated cigar-shaped trap potential has limitations and is ineffective, especially in real space.

We next consider large trap potentials using the same values for $N$, $g/d^2$, and $m$ as those given in the previous paragraph. For $\omega=11\pi \times 10^2$ Hz, $\beta_0{\rm =1.0}$, and $T=1.5T_c^0$, we have aspect ratios $\alpha^{-3/2}\simeq 0.912$ in the momentum distribution and $\beta^{3/2}\simeq 1.091$ in the spatial distribution. For $\omega=30\pi \times 10^2$ Hz, $\beta_0{\rm =1.0}$ and $T=2.0T_c^0$, we have $\alpha^{-3/2}\simeq 0.930$ and $\beta^{3/2}\simeq 1.072$. Thus, the deformation effects can be observed by using a high trap frequency rather than by using a high trap aspect ratio.

\section{Conclusion}

We have used a variational method to study the equilibrium properties of a dipolar gas at finite temperatures. For a Fermi gas at zero temperature, the anisotropic nature of the dipolar interaction leads to deformations in momentum and real space, and the partial attraction of the interaction causes the gas to collapse. In addition, the dipolar Fermi gas becomes compressed in momentum space as the electric dipole moment increases. We found that the deformations in both momentum and real space can be observed in the high-temperature regime with a large electric dipole moment and a high trap frequency. In addition, the stable region expands at finite temperatures. These results will be useful when cooling polar molecules in experiments.

By studying a Bose gas, we find that the different statistics of Fermi and Bose gases gives rise to different signs for the Fock exchange energy as well as the emergence of the interatomic interaction in dipolar Bose gases. The different signs of the Fock exchange energy causes opposite properties: the momentum distribution is elongated in the dipole direction in a dipolar Fermi gas, whereas the momentum distribution is stretched perpendicular to the dipole direction in a dipolar Bose gas.

In addition, we found that the local minimum of the free energy always disappears when $\lambda$ exceeds a critical value for both Fermi and Bose gases. This implies that in both cases the instability first occurs in real space. However, the different statistics mainly arise from the sign of the Fock exchange energy, which indicates that the difference mainly arises in momentum space. Consequently, there are few differences in the critical interaction and the critical temperature as a function of the trap aspect ratio for Fermi and Bose gases. On the other hand, there is a clear difference in the critical temperature as a function of the interaction for the two systems. The dipolar interaction changes from being attractive to being repulsive on decreasing the trap aspect ratio. On the other hand, the repulsive interatomic interaction keeps the system stable, and the attractive interatomic interaction causes the system to become unstable. For this reason, decreasing the ratio of the interatomic interaction to the dipolar interaction causes the instability to become independent of the trap aspect ratio for a Bose gas.

Our variational formalism can be extended to study the dynamics of dipolar gases at finite temperatures, such as the expansion dynamics and collective oscillations. In particular, expansion dynamics is important since it can provide direct information about deformation of the momentum distribution. This type of dynamics is also useful in studying the different statistics, which is one of the main results of this study. In addition, it may be interesting to study the expansion dynamics of a dipolar Bose--Fermi mixture. We are also interested in the equilibrium and dynamics below the Bose--Einstein transition temperature for dipolar Bose gases. We note that the relation between the real-space distribution and the momentum-space distribution of BECs is described by the Fourier transform of the macroscopic wave function, but it is not the case for a thermal cloud gas. Therefore, fascinating properties are expected below the Bose--Einstein transition temperature.

In the future, we intend to investigate the temperature dependence of the expansion dynamics and collective mode in the presence of the dipolar interaction. In addition, we will study the equilibrium and dynamics of a dipolar Bose--Fermi mixture and dipolar Bose gases below the Bose--Einstein transition temperature.
We hope this study will stimulate further experiments on dipolar gases at finite temperatures.

\section{Acknowledgement}

Y. E. is supported by a Grant-in-Aid for JSPS fellows and T. N. is supported by a Grant-in-Aid for scientific research from JSPS.
\appendix

\section{Deformation function}\label{AppendixA}

\begin{figure}[h]
  \begin{center}
    \begin{tabular}{cc}
      \scalebox{0.5}[0.5]{\includegraphics{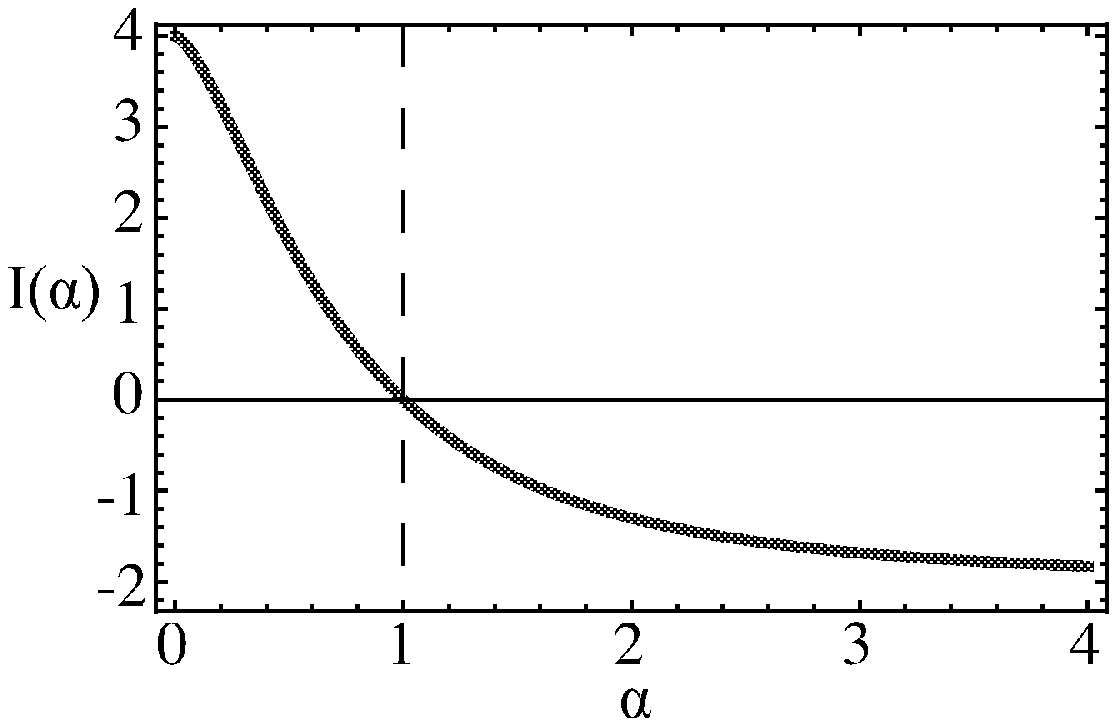}} &
      \scalebox{0.5}[0.5]{\includegraphics{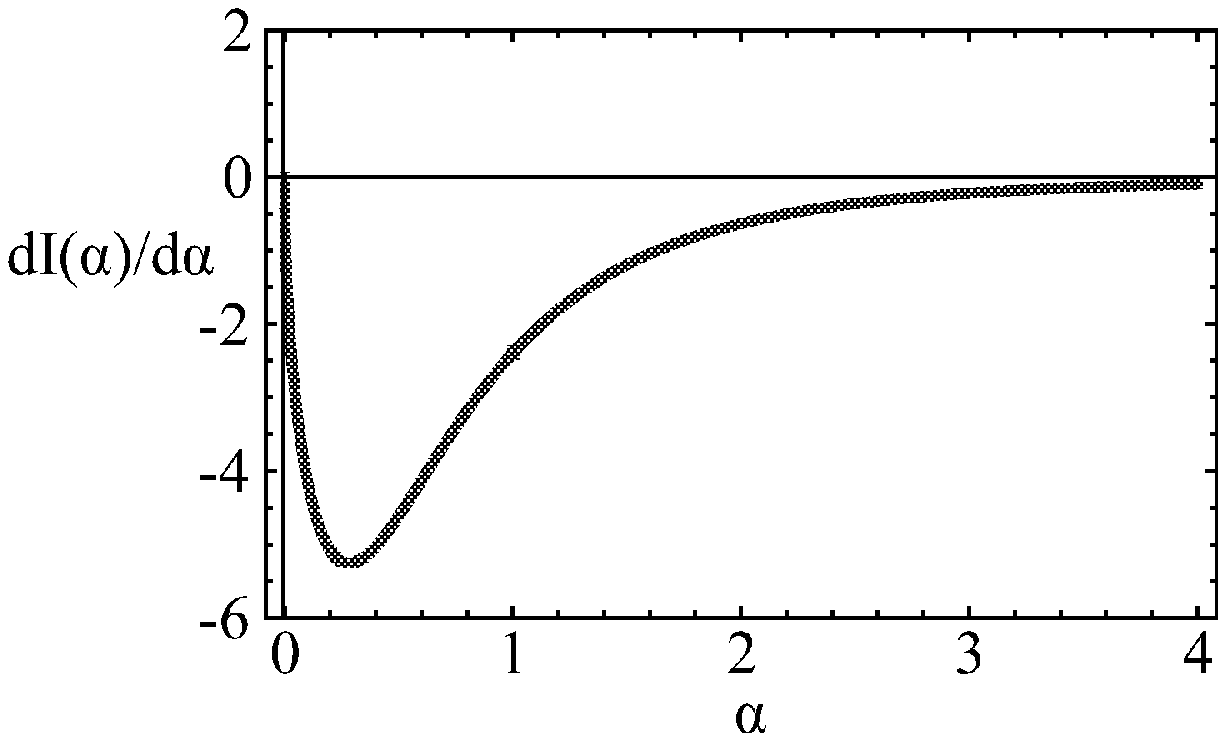}}
    \end{tabular}
    \caption{Deformation function $I\left(\alpha\right)$ and the derivative of $I\left(\alpha\right)$ as a function of $\alpha$.}
    \label{fig:DeformationFunction}
  \end{center}
\end{figure}

In this appendix, we consider the deformation function~(\ref{deformation_function}), which determines the properties of the Fock exchange energy and the Hartree direct energy. Rewriting the deformation function, one can derive the analytical expression~\cite{erratum1}:
\begin{eqnarray}
I\left(\alpha\right)&=&\int_0^\pi d\Theta \sin{\Theta}\left(\frac{3\cos^2\Theta}{\alpha^3\sin^2\Theta+\cos^2\Theta}-1\right)\nonumber \\
	&=&\left\{
\begin{array}{l}
	\displaystyle \frac{6}{1-\alpha^3}\left[ 1-\sqrt{\frac{\alpha^3}{1-\alpha^3}}\arctan\left(\sqrt{\frac{1-\alpha^3}{\alpha^3}}\right) \right]-2 \ \ \ \ \ \ \ \left(\alpha<1\right)\\
	\displaystyle 0 \ \ \ \ \ \ \ \ \ \ \ \ \ \ \ \ \ \ \ \ \ \ \ \ \ \ \ \ \ \ \ \ \ \ \ \ \ \ \ \ \ \ \ \ \ \ \ \ \ \ \ \ \ \ \ \ \ \ \ \ \ \ \ \ \ \ \ \left(\alpha=1\right) \\
	\displaystyle -\frac{6}{\alpha^3-1}\left[ 1-\frac{1}{2}\sqrt{\frac{\alpha^3}{\alpha^3-1}}\log\frac{\sqrt{\alpha^3}+\sqrt{\alpha^3-1}}{\sqrt{\alpha^3}-\sqrt{\alpha^3-1}} \right]-2 \ \ \ \left(\alpha>1\right)
	\end{array}
	\right.
\end{eqnarray} 
This formula indicates that $I\left(\alpha\right)$ is a monotonically decreasing function of $\alpha$ with $-2\leq I\left(\alpha\right)\leq 4$; it is positive for $\alpha<1$, passes through zero at $\alpha=1$, and is negative for $\alpha>1$~\cite{Miyakawa2008,Sogo2009}. Figure~\ref{fig:DeformationFunction} shows the deformation function and its derivative.

\end{document}